# A Distributed Spatial Data Warehouse for AIS Data

Alex S. Klitgaard    Lau E. Josefsen    Mikael V. Mikkelsen    Kristian Torp
Aalborg University
Dept. of Computer Science
Aalborg, Denmark
alex.klitgaard@gmail.com,lau@josefsens.dk,mvmi@cs.aau.dk,torp@cs.aau.dk

**Abstract**

AIS data from ships is excellent for analyzing single-ship movements and monitoring all ships within a specific area. However, the AIS data needs to be cleaned, processed, and stored before being usable. This paper presents a system consisting of an efficient and modular ETL process for loading AIS data, as well as a distributed spatial data warehouse storing the trajectories of ships. To efficiently analyze a large set of ships, a raster approach to querying the AIS data is proposed. A spatially partitioned data warehouse with a granularized cell representation and heatmap presentation is designed, developed, and evaluated. Currently the data warehouse stores 312 million kilometers of ship trajectories and more than 8 billion rows in the largest table. It is found that searching the cell representation is faster than searching the trajectory representation. Further, we show that the spatially divided shards enable a consistently good scale-up for both cell and heatmap analytics in large areas, ranging between 354% to 1164% with a 5x increase in workers

## 1 Introduction

The automatic identification system (AIS) data collected from ships shows promise for analytics [16]. For example, the explosions at the Nord Stream underwater gas lines on the $26^{th}$ of September 2022 [2] show how AIS can be used to determine which ships were near the gas lines at the time of the explosion

Although AIS data show great promise for analytics, sufficient care must be taken to overcome its inherent limitations [16]. One of the inherent limitations of AIS data is that it is dirty, as the protocol was originally designed to help increase safety at sea. Therefore, before using AIS data for analytics, it is necessary to clean the data based on the knowledge about the ship domain. Danish AIS data have been stored for almost two decades resulting in a large amount of available AIS data for analytics, such as traffic pattern mining [34] and draught maps [28].

The large volume of available AIS data makes storing and processing the data on a single machine infeasible. Therefore, systems proposed for efficient handling of trajectory data, such as [9, 35], use a distributed engine to store the data on a cluster of workers.

The DIPAAL platform consists of a modular ETL process responsible for cleaning and transforming the AIS data before loading the data into a PostgreSQL-based data warehouse (DW). The DW supports a trajectory and a cell representation of the AIS data as shown in Figure 1.

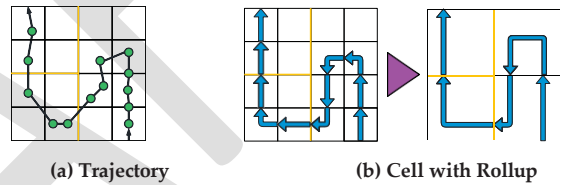

(a) Trajectory         (b) Cell with Rollup

**Figure 1: AIS data in DIPAAL with Spatial Divisions**

Trajectories are chronologically ordered sets of AIS points, as shown in Figure 1a, where the green circles are individual AIS data points. Whereas cells are spatial areas that aggregate trajectory information, as shown in Figure 1b. The cell representation consists of a hierarchy that stores the resulting cells at multiple granularities. The Citus PostgreSQL extension enables distribution by sharding the data in the DIPAAL DW across a cluster of workers. A novel spatial distribution approach is used to enable spatial aggregations on the cell and heatmap representations. This approach creates spatial divisions that divide the cell and heatmap data across all available workers. These spatial divisions are created using a kd-tree approach and are based on an entire year's worth of AIS data.

An example of spatial divisions is the yellow lines in Figure 1 which show a large division on the right and two smaller divisions on the left. It is evident from Figures 1a and 1b that the cell representation is more suited for spatial distribution than the trajectory representation, as trajectories can span multiple divisions, whereas cells fit within a single division.

This paper explores how a kd-tree-based spatial distribution approach impacts the query runtime of different query types and the utilization of the underlying hardware. Additionally, how well the spatial distribution approach scales when going from a one- to five-worker setup is evaluated for queries with areas of different sizes.

The contributions of this paper are:

- Implementation of a spatial distribution approach that reduces query runtime on the cell representation by ensuring spatial data locality on the workers and enabling push down of spatial aggregation to the workers.
- On-demand distributed heatmap creation for analytics
- Extensive evaluation of the DIPAAL platform which offers insights into the query runtime performance of the platform and how well it scales from a one to a five worker setups.
- Show the feasibility of the design of DIPAAL by loading, storing, and performing analytics on five years of AIS data.

Currently, 5 years are loaded into the DW, consisting of 414 million kilometers of ship trajectories between 13.9 million trajectories over a period of 1 111 143 days. A total of 57 238 unique ship IDs (MMSIs) are observed. The largest relation contains 10 billion rows.

The rest of the paper is structured as follows. The definitions used and the architecture of DIPAAL are described in Section 2. Section 3 explains the ETL design and the process used to load data. The structure and design choices behind the data warehouse are covered in Section 4. Section 5 discusses approaches to spatial divisions. Section 6 covers implementation. Section 7 evaluates the efficiency of DIPAAL. Related work is explained in Section 8. Finally, Section 9 concludes the paper.

## 2 Architecture

This section describes the architecture of DIPAAL. First, the key definitions are given. This is followed by looking at the overall system architecture. Finally, details related to the design of the data warehouse are presented.

### 2.1 Definitions

**Definition 1.** *Point: A point $p = \{t, lng, lat, MMSI, ais\}$, where $t$ is the timestamp the point was created, $lng$ and $lat$ are abbreviations for the longitude and latitude coordinates. MMSI refers to the Maritime Mobile Service Identity of the transmitting ship, and ais are additional optional AIS attributes, e.g., draught and ship type.*

MMSI is a 9-digit identification number that is mandatory to use for identification [12]. The additional optional attributes *ais*, in Definition 1, are the attributes contained within the AIS message data related to the ship. Examples include the speed over ground (SOG), destination, heading, and navigational status, as explained by Bereta et al. [12].

**Definition 2.** *Trajectory: A trajectory $T$ is a sequence of points sorted on their timestamp in ascending order $T = \{p_1, p_2, ..., p_n\}$, where $n \geq 2 \land p_1.t < p_2.t < ... < p_n.t$.*

A trajectory must consist of at least two points in DIPAAL. Otherwise, it cannot express movement over time. The points of the trajectory must be temporally ordered.

Definitions 1 and 2 are data source independent and applied to most, if not all, AIS datasets.

As both the sharding and partitioning techniques are used in DIPAAL, it is necessary to define the difference between sharding and partitioning clearly.

**Definition 3.** *Shard: Relations are divided into shards to scale databases horizontally. Each shard is a subset of the parent relation and may be located on different workers in a distributed database. This is called sharding [30].*

**Definition 4.** *Partition: A partition refers to splitting a logical relation into multiple physical smaller relations on the same worker [19].*

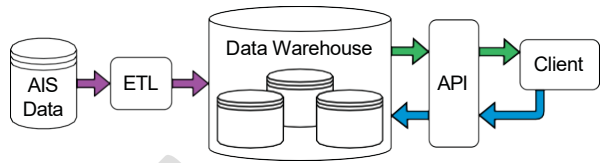

**Figure 2: The DIPAAL Architecture**

### 2.2 Platform

Figure 2 shows the architecture of DIPAAL, where purple arrows depict data processing, green arrows depict requests for data, and blue arrows depict the responses. Cylinders indicate data storage, and boxes are software components.

To improve data quality, a set of cleaning rules are defined. The ETL process follows the rules defined by Nielsen et al. [28]. These cleaning rules are similar to the work of Graser [18].

The data warehouse stores the AIS data processed by the ETL process. This results in multiple representations being available. As defined in Definition 2, the coarse grains are cells and the finest granularity is trajectories, as shown in Figure 1.

Power users can query the data warehouse directly. Other users can send REST API requests to pre-optimized endpoints. The API is available at https://dipaal.dk. Examples of the images and rasters produced by the API are available at https://dipaal.github.io

### 2.3 Data Warehouse

The underlying storage engine must support spatio-temporal data and distributed storage to create an efficient query platform for large-scale trajectory data. PostgreSQL supports spatio-temporal data and distributed storage through extensions. DIPAAL uses the extensions Citus [3] for distribution and PostGIS [5] plus MobilityDB [4] for spatio-temporal supports.

*Distributed Storage Support* Citus enables distribution on PostgreSQL by sharding relations (Definition 3) and storing these shards across a cluster of workers. Citus supports co-located relations [14], meaning rows with the same value in their distribution attribute are located on the same worker. Co-location increases data locality and enables local joins [14], significantly reducing query runtime. In contrast, non-co-located joins must transfer data between workers to perform a join.

Distribution of the data also enables the workers in the Citus cluster to share the workload by each computing part of a query result from its local data, which are combined and presented to the user [13].

*Spatio-temporal Support* PostGIS extends PostgreSQL with a range of spatial data types, operators, and functions. PostGIS is used for raster operations, to enable the heatmap functionality of DIPAAL, as well as to provide spatial join conditions.

MobilityDB builds on top of the PostGIS extension and adds support for temporal and spatio-temporal data by lifting the operations and functions provided by PostgreSQL and PostGIS [10]. By defining multiple type constructors, MobilityDB enables working with both discrete spatio-temporal data, as well as providing Moving Object Database (MOD) capabilities [10].

## 3 ETL Design

The ETL process consists of five interchangeable and self-contained modules to facilitate the separation of concerns and improve modularity. The process consists of five main steps where the first three presented are data warehouse independent. These parts can be executed in parallel on multiple machines, with no coordination. The individual modules are described in detail below. Here we describe a base use case with a synchronous execution of the following steps.

*File Download Module* This module provides a thin integration that enables the ETL process to find, download, and prepare the raw AIS source files automatically. An example is the unzipping of archived files.

*Data Cleaning Module* This module removes dirty AIS data by enforcing the domain-specific cleaning rules, defined by [18, 28]. The main parts done are removing ships with unnatural dimensions, an incorrect MMSI, or a position on land.

*Trajectory Construction Module* The *trajectory construction module* is responsible for constructing the trajectory stored in the data warehouse, see Definition 2.

The module creates trajectories from a temporally ordered AIS data set. To handle each ship in isolation, the input data are grouped based on the ship's MMSI value. MMSI is a mostly unique identifier, as it uniquely identifies a ship at a given time, but it may change under certain circumstances [1, 12].

The trajectories are classified as either stopped or moving, enabling analysis to focus on either classification.

To construct trajectories, the *trajectory construction module* iterates over the temporally ordered AIS data set and examines the points pairwise. From this examination, the module determines which set of points represents moving and stopped ships, creating trajectories for each. The pairwise examination is further used to improve the data quality by removing spatio-temporal outliers, as defined in [28].

*Rollup Module* The *rollup module* executes the rollup queries that transform trajectory data into cell-related data, as seen in Figure 1. The work is trivially distributed and parallelized, as Citus automatically creates a distributed query plan.

## 4 Data Warehouse Design

The design of DIPAAL is based on a collection of sources and recommendations and serves as a proposal for efficiently distributing analytics of large-scale moving object data.

The design is partially based on the recommendations of Kimball and Ross [20]. It is partially based since Kimball and Ross do not make recommendations for some aspects of the spatio-temporal and moving-object world, such as spatial references and representations. For these aspects, the design is based on the recommendations made by Vaisman and Zimányi [32] and from consultations with domain experts. The overall design is based on a star schema with an RDBMS as the physical foundation, as recommended in [20].

Figure 3 shows the complete data warehouse design. It comprises six fact tables that individually are part of a partially snowflaked star schema with conformed dimensions except for the time and date dimensions which are role-playing dimensions. The complete schema is a fact constellation schema [27].

Each attribute in the fact constellation schema is depicted using three sections, where the leftmost section specifies if the attribute is part of the primary key (PK) and/or a foreign key (FK). The middle section is the attribute's name, and the rightmost section is the data type. Attributes that are both part of the PK and a FK are represented with a green color, while attributes that are only part of a PK are represented with a blue color. If a PK attribute is the distribution attribute, it has a purple gradient. The distribution attribute determines which Citus shard the row lies in. Each arrow points from a FK to the attribute it references, as seen in Figure 3.

Arrows to dim_spatial_division, dim_date, and dim_time are not shown but are implicit for all attributes named with *division_id*, *date_id* and *time_id*. **Tgeompoint**, **stbox**, and **tfloat** are types from MobilityDB, where **tgeompoint** represents temporal geometric points, **stbox** represents a spatio-temporal box, and **tfloat** represents temporal floats [4].

Four granularities 5000m, 1000m, 200m, and 50m are chosen for the cell representation. This is the result of collaboration with domain experts and their current use of nautical maps.

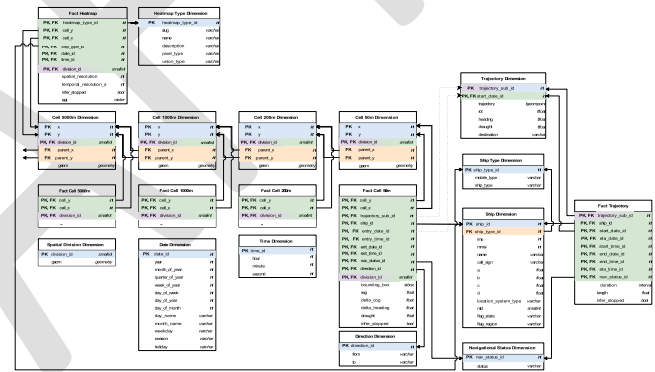

**Figure 3: Fact constellation schema for the DW.**

### 4.1 Facts

Kimball and Ross recommend that the fact table foundation builds on atomic events [20]. However, the data foundation of each row in the fact table fact_trajectory is a trajectory. This granularity is used for fact_trajectory, as no information is lost by aggregating AIS points into trajectories; see Definitions 1 and 2.

Fact_trajectory contains measures that describe a ship's trajectory, as defined in Definition 2 and seen in Figure 1a. The measures duration and length represent the duration of the trajectory and its length in meters, respectively. These are additive measures [20]. Lastly, the measure infer_stopped is a Boolean where true means the trajectory represents a non-moving ship, for example, an anchored ship. False if for a trajectory from a moving ship.

The four cell fact relations are aggregate fact tables based on fact_trajectory, as shown in Figure 1, and contain measures for events within defined cells. These events describe the movement of a ship between entering and exiting a cell. The `SOG` averages the ship's SOG during the cell event. `Delta course over ground` (COG) and `heading` are measures of how much the ship's course and heading changed inside a cell.

The `fact_heatmap` relation contains entries defining a raster for a 5 by 5km cell. The 5 by 5km raster contains the finer granularities by storing multiple pixels in the raster. For example, the 50m resolution raster is a 5 by 5km raster containing 10,000 pixel values. This reduces query runtime because fewer rasters are aggregated.

*Rast*, is an advanced [32] and semi-additive measure, as it cannot be aggregated across different heatmap types or granularities. The `fact_heatmap` relation serves as an intermediary pre-aggregate of the four cell fact relations, where entries of the relation can be aggregated further to create heatmaps. Pre-aggregation minimizes the computation needed at query time, while preserving the ability to create parameterized heatmaps.

The `fact_heatmap` relation also contains measures for the spatial and temporal resolution in spatial reference units, which for the Danish data foundation is meters from EPSG:3034, and seconds, respectively. These measures serve to enable storing multiple resolutions. DIPAAL currently pre-aggregates heatmaps with spatial resolution of the four cell granularities, and one day as the temporal resolution, as per request by domain experts.

### 4.2 Dimensions

Both `fact_trajectory` and the four cell fact relations reference the dimension `dim_trajectory`. This dimension is responsible for storing the variable length fields of a trajectory. It stores a **tgeompoint**, representing the trajectory, but also **tfloat**s for the *rate of turn* (ROT), *heading*, and *draught*. Lastly, it has a **varchar** attribute representing a trajectory's *destination*.

The `dim_heatmap_type` stores metadata about different heatmap types, such as how to aggregate the rasters, and its name and description. At the time of writing, DIPAAL, pre-aggregates five heatmap types; count of ships crossing a cell, accumulated time spent in a cell, average delta change in heading in a cell, average delta change in COG in a cell, and minimum draught in a cell. A two-band raster is used to support aggregating average values, with one band being the sum of the value and the other being the count. The `dim_heatmap_type` enables the possibility of defining new heatmap types dynamically without any structural changes to the DW, thus only requiring trivial ETL changes.

The four cell dimensions all store a two-dimensional index of cells and their physical geometries using the PostGIS **geometry** datatype. The four cell dimensions use the same hierarchy and granularity levels as the four cell facts. The four cell dimensions are linked in a snowflake hierarchy, with smaller cells referencing their larger parent cell.

A snowflake approach is chosen to avoid the data explosion, which occurs if the hierarchy is represented as a denormalized star schema hierarchy. Denormalizing the hierarchy with 5000m and 50m cells results in the 5000m granularity being replicated 10,000 times. Furthermore, a snowflake approach trivializes adjusting the hierarchy in the future.

`Dim_spatial_division` is a non-changing fixed dimension. It contains two attributes, consisting of an id and a geometry. This dimension is used for every relation that is spatially distributed. The spatial distribution scheme is described in depth in Section 5.

### 4.3 Distributed Relations and Attributes

The dimensions `dim_nav_status`, `dim_direction`, `dim_spatial_division`, and `dim_time` are fixed size. The `dim_date` relation grows on average by one row for each loaded day, and `dim_ship` grows by the number of new ships every day but is expected to grow by less than a few hundred rows per day. All these relations are configured as reference relations, meaning all the data in these relations are replicated to all workers in the Citus cluster to ensure data locality [14].

The relations `fact_trajectory`, `dim_trajectory`, the four cell facts, and `fact_heatmap` are expected to grow the fastest. For this reason, these are distributed. The four cell fact and `fact_heatmap` relations are distributed spatially using the `dim_spatial_division` dimension, described in detail in Section 5. The four cell dimension relations are spatially distributed together with the four cell fact relations, ensuring data locality.

For the `fact_trajectory` and `dim_trajectory` relations to be distributed spatially, it requires the trajectories to be split into segments that fit inside the divisions defined in `dim_spatial_division`. Splitting the trajectories requires trajectory reconstruction if the complete trajectory is needed for a query, involving cross-node reconstruction and repartitioning. It is thus chosen to instead use random hash partitioning for the trajectories. This design choice means that to create cell facts from the trajectories, a repartitioning has to occur, as the cell relations are not co-located with the trajectory relations. This issue is discussed in detail in Section 6.2.

## 5 Spatial Distribution

DIPAAL can produce a variety of heatmaps. The production of heatmaps is a computationally heavy task, scaling with the temporal and spatial spans. Therefore, the distribution of such tasks and the distribution method used are important.

### 5.1 Motivation

To facilitate efficient distributed calculation of heatmaps, all values of a pixel must be located in the same shard, as the summation can then be pushed down to the workers. Rasters of the same colors indicate rasters of the same area but may represent, for example, a different time, ship type, or mobile type. An example of two rasters of the same color is one for the ship type *Cargo*, and one for the ship type *Passenger*. The notion *area* refers to the area covered by the raster.

Aggregating the rasters on the coordinator is expensive in network overhead and limits the system to the computational power of the coordinator. It is, therefore, not horizontally scalable. The spatial distribution guarantees that all rasters for a given area are located on the same worker. Thus, the workers perform the raster aggregation for each area and only transmit one raster per area to the coordinator. The coordinator performs less work; the only operation it has to do is align the areas into a large raster.

To horizontally scale spatial aggregates, the cell facts and dimensions are spatially co-located with the heatmaps.

Quad-tree and kd-tree-based spatial distributions for the spatial domain of DIPAAL are seen in Figure 4, which is explained next.

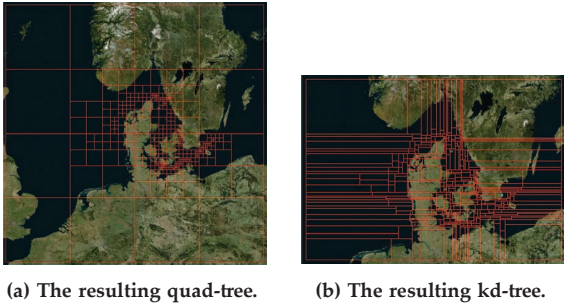

**(a) The resulting quad-tree.**  **(b) The resulting kd-tree.**

**Figure 4: Quad-tree and kd-tree Spatial Division**

## 5.2 Spatial Distribution Approach

Spatial distribution is required to scale spatial aggregations, such as heatmaps. The next step is determining the approach to create the spatial distributions by dividing the spatial domain.

As DIPAAL has a maximum cell granularity of 5km, no spatial division can have a side length not divisible by 5km, and no division can thus be smaller than 5 by 5km.

The spatial distribution aims to create divisions such that the counts of cell facts are balanced between the shards. This is because the four cell fact relations are by far the largest. Creating equal-size shards ensures that both the load, when querying the cell relations, and the storage requirements for each worker are balanced. As the count of the cell facts of different granularity is proportional, the count of 5000m cell facts is used as the foundation of the spatial distribution.

Two division approaches are considered for use in DIPAAL. The first is a region quad-tree approach [31], and the second is a kd-tree approach [11]. R-trees can also be used for spatial division, but since they do not spatially cover the domain, it is discarded as an option. Using R-trees for spatial division was explored by Li et al. [24], where the non-covering property resulted in a work-around that ended up performing poorly compared to quad- and kd-trees.

For the quad-tree construction, a global spatial domain is chosen by finding the shortest side lengths that surround the global spatial domain of the data while having side lengths, $s$, be $s \in 2^n * 5000m$ where $n$ is a natural number. The side length must be 5,000m multiplied by a two exponent, as each quad-tree split divides the cell's side lengths in half. This multiplication ensures each cell, even after $n$ splits, is divisible by 5km. Consequently, $n$ defines the maximum depth of the quad-tree. For each split, the division with the highest number of cells is chosen until it reaches a maximum depth, such that splitting again results in divisions smaller than 5 by 5km.

A similar approach is taken for the kd-tree construction. The input to the function is the global spatial domain and a limit on the number of divisions created.

Both the quad- and kd-tree approach have a parameter defining the maximum number of divisions created. 400 divisions are chosen as the maximum number of divisions based on experimentation in the year 2021. This experimentation shows that creating additional divisions requires splitting divisions into sizes less than 5km or introducing a higher imbalance of the divisions.

The resulting spatial divisions of both approaches do not change over time, i.e., they are static. The static property enables the push down of spatial aggregation to the workers, regardless of the query's temporal range. The consequence of static division is the possibility of divisions becoming skewed as traffic patterns evolve or new AIS receivers are installed. Data deviation over time is evaluated in Section 7.4.

The resulting quad- and kd-tree are shown in Figure 4, with the quad-tree approach in Figure 4a and the kd-tree approach in Figure 4b.

The data deviation is compared using the measures of Standard Deviation (SD) [26] and Coefficient of Variance (CV) [6]. For both metrics, a lower value indicates better-balanced shards. The 2021 data contain 9 million trajectories, traveling a total of 85 million kilometers by 28k unique ships.

After constructing the quad- and kd-trees on AIS data for 2021, the SD and CV are measured. For the quad-tree, the SD is 81,192, and the CV is 106%. For the kd-tree, the SD is 31,814, and the CV is 41%. Kd-tree-based divisions are better balanced and are used as the distribution approach for DIPAAL. This result is explained by the unevenness of the data foundation, which is due to the nature of navigation on water that mostly follows defined sea lanes, resulting in high- and low-traffic areas. The dynamic nature of the kd-tree approach better captures this unevenness compared with the more rigid quad-tree approach, and thus, these results may not apply to other spatial domains or datasets.

## 6 Implementation

This section presents the implementation details of DIPAAL. The source code of DIPAAL is found at https://github.com/DIPAAL/etl.

### 6.1 Line Simplification

During the implementation of the ETL process, it is observed that constructed trajectories contain points with high spatio-temporal similarity, especially for stopped ships. Each trajectory is stored as a MobilityDB sequence of points. MobilityDB sequences imply a linear interpolation between points with consecutive timestamps, allowing the spatial position of a ship to be inferred at any given time within a trajectory. Therefore, removing data points that contribute little to the precision of the trajectory can thus reduce the query runtime of DIPAAL without significant precision loss.

To increase the performance of DIPAAL, line simplification [17] is performed for each trajectory in the dim_trajectory relation. DIPAAL performs line simplification with an error bound of 10 meters, reducing the number of points stored by a factor of six. The Douglas–Peucker algorithm [15] with Synchronized Euclidean Distance [29] (SED) is used for line simplification through the douglasPeuckerSimplify MobilityDB function. SED differs from the Euclidean distance as it also considers the temporal aspect [29]. As a consequence of the simplification process, the trajectories are introduced to an error of 10 meters. Domain experts approve of introducing this error, as the benefits of a six-fold data reduction outweigh the minor errors introduced.

### 6.2 Rollup

The *rollup module* is responsible for applying rollup, resulting in aggregated representations of the trajectory data. The functionality of this module is implemented as a set of SQL queries, as this module only interacts with the data warehouse.

*Cell Rollup* Trajectories are aggregated to populate the `fact_cell` relations for each cell granularity. Experiments show that populating `fact_cell` by directly joining `dim_cell` to the geometry representing the trajectory in `dim_trajectory`, results in poor performance. This is due to a combination of PostGIS functions necessary for such a join, `ST_Crossing`, `ST_Contains`, and `ST_Touches`, suffering from an inefficient index filtering when the trajectories' Minimum Bounding Rectangle (MBR) is large.

*Heatmap Rollup* Cell facts are aggregated into rasters to populate the `fact_heatmap` relation with heatmaps.

The efficient creation of heatmaps is facilitated by the spatial distribution of the `fact_cell` relations, as explained in Section 5, as it enforces data locality.

The spatial distribution allows each worker to locally create a section of the total heatmap, a raster. These rasters populate the `fact_heatmap` relation and are created for each granularity of `fact_cell` in the data warehouse. Additionally, different types of rasters are created for each granularity, each aggregating a particular type of information from `fact_cell`.

## 7 Evaluation

This section evaluates the performance of a range of queries on DIPAAL and the data skewness of the spatial distribution approach with new and old data. The evaluation focuses on query runtime in a single-user scenario and thus does not provide any evaluation of a concurrent multi-user scenario.

### 7.1 Hardware/Software

The DW cluster of DIPAAL is a Citus cluster and consists of six identical machines (AMD Ryzen 7 5800x, 32GB RAM, 2 TB Disk) in a setup of one coordinator and five workers. The machines are connected through Ethernet on a gigabit connection.

Each machine runs a containerized instance of PostgreSQL 15.3 with the extensions PostGIS 3.3.2, MobilityDB 1.1.0, and Citus 11.3.0 in Kubernetes [21]. The OS is Ubuntu 22.04 with Python 3.11.

The dataset, which DIPAAL is evaluated upon, is the publicly available AIS dataset published by DMA. The year 2021 is loaded for all benchmarks unless otherwise stated in the benchmark section.

### 7.2 Runtime Evaluation Metrics and Procedure

Evaluations that measure runtime all use two metrics; runtime and average Worker Idle Fraction (WIF)

The runtime is the duration from sending the query to the DW to receiving a complete result. WIF is defined as the fraction of the runtime a worker spends idling. A high WIF indicates that the computational power of the worker is wasted. WIF is calculated as defined in Equation (1)

$$\text{WIF}_j = \frac{\max(\text{Worker Time}) - \text{Worker Time}_j}{\max(\text{Worker Time})} \quad (1)$$

In Equation (1), Worker Time$_j$ is the time that Worker $j$ took to finish all shards located on that worker and is defined in Equation (2)

$$\text{Worker Time}_j = \max_{i \in \text{Worker Shards}_j} \text{Shard Time}_i \quad (2)$$

Average WIF has a lower bound of 0%, but an upper bound of $1 - \frac{1}{|workers|}$. The upper bound is due to a single worker always being non-idling. For this reason, the average WIF is normalized to a range from 0% to 100%, as seen in Equation (3)

$$\text{Average WIF} = \frac{\sum \text{WIF}}{|workers|} * \frac{1}{1 - \frac{1}{|workers|}} \quad (3)$$

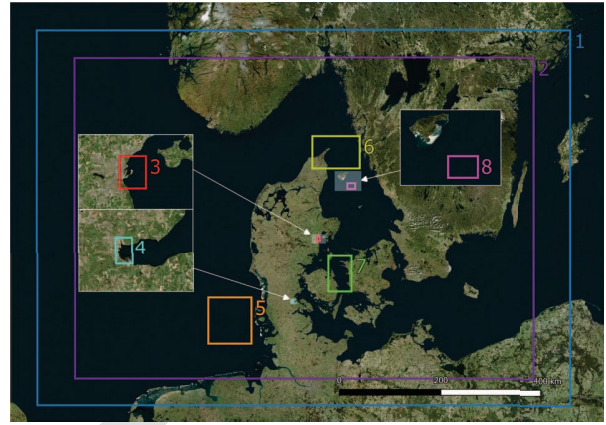

**Figure 5: The Eight Benchmark Areas**

Each query is evaluated through ten iterations to remove outliers from the result. The metrics are then aggregated using a trimmed mean, by removing the two lowest and two highest results, before taking the mean.

To ensure consistent results, all PostgreSQL instances in the cluster are restarted, and the OS file cache is cleared before each query iteration. The cache is then pre-warmed by random queries.

### 7.3 Heatmap Evaluation

This section covers the heatmap performance evaluation of DIPAAL. The heatmap benchmark serves two purposes; To find the best-performing physical storage scheme and to evaluate the scale up when comparing a one-worker setup to a five-worker setup.

Three physical storage schemes are evaluated; row-based storage, Citus Columnar, and partitioned Citus Columnar. The partitioned Citus Columnar storage scheme uses Citus Columnar but is partitioned on heatmap type, resolution, and month, instead of global monthly partitions. The Partitioned Citus Columnar scheme aims to reduce the Columnar runtime, by storing data that are likely to be queried together in the same stripe.

These three physical storage schemes are evaluated on the five-worker setup. The best performing physical storage scheme is evaluated on the one-worker setup to compare the scale up.

The evaluation of a configuration consists of a series of 36 queries. The 36 queries are the cross product of the four cell granularities, three temporal spans, and three spatial areas, and are seen in Figure 6. The three temporal spans are 1 day (February 28th, 2021), 30 days (January 26th through February 24th, 2021), and 1 year (2021). The three spatial areas are seen in Figure 5, with Aarhus Harbour ($46.78 km^2$) in red (3), The Great Belt ($3\,071 km^2$) in green (7), and the spatial domain of DIPAAL ($725\,725 km^2$) in blue (2).

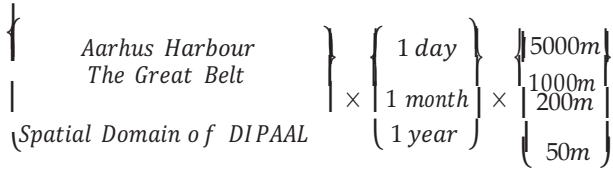

Figure 6: The 36 Heatmap Configurations

The results of the evaluation of heatmaps are seen in Table 1. The 1-day queries are not included as they are always below one second and are thus uninteresting.

*Evaluation of Physical Storage Scheme* In Table 1, the fastest configuration for each query is marked in bold. It is noticeable that the five-worker setup with row-based storage is the fastest at the majority of the queries, besides the queries with a runtime at or below one second, where the difference is insignificant.

As it is more important to reduce the runtime of the slow queries, the row-based physical storage scheme is chosen as the physical storage scheme used in DIPAAL. The row-based physical storage scheme is chosen as it has the best performance across all queries with a significant runtime.

*Evaluation of Scale Up* The row-based physical storage scheme is applied to a one-worker setup and compared to the row-based physical storage scheme of a five-worker setup. This result is seen in the right-most column in Table 1. The scale up is measured in percent, with 100% indicating the same runtime and 200% indicating a runtime two times faster on the five-worker setup.

It is seen in Table 1, that the scale up varies significantly. This variation is expected as smaller areas only engage a few shards, which is also indicated by the high WIF in the *Aarhus Harbour* query. It is seen that the WIF is lower the larger the area queried, as more shards are engaged in calculating the results. In the *Aarhus Harbour* area, five shards are engaged; in the *The Great Belt* area, 15 shards In the it full spatial domain, all 400 shards.

On the queries with a lower runtime, i.e., the queries in the areas of Aarhus Harbour and the Great Belt, on the five-worker setup with row configuration, the WIF is consistently lower for 200m heatmaps than 50m. Likewise, it is observed that the scale-up is worse for 50m than 200m heatmaps. By investigating the explain analyze dump, it is clear this is due to a shard imbalance. For example, in the one-year Great Belt query, 14 shards are involved in the calculation.

The largest shard holds 2.4GB of raster data, while the next-largest holds 1.9GB of raster data, i.e., a 26% size difference. By investigating the query plan and the individual shard timings, it is observed that the runtime of the spatial aggregation of the rasters is linearly proportional to the sum of pixels in the shard.

## 7.4 Spatial Distribution Skew

The spatial distribution skew benchmark differs from the other benchmarks by evaluating how skewed the spatial distribution becomes over time rather than measuring runtime.

To measure the skewness, Standard Deviation (SD), as well as Coefficient of Variance (CV) are used, as introduced in Section 5, where lower values for both metrics indicate better-balanced shards. The metrics are measured for each of the four cell granularities regarding the count of cell facts and the sum of the size of heatmap rasters.

In Table 2, it is seen that the CV of both the count of cell facts and the size of heatmap rasters for all granularities are higher in 2011 and 2022 than in 2021. It is also seen that 2022 is much closer to the 2021 measurement than 2011, which is to be expected, as traffic patterns are more likely to have changed over ten years versus one year. The evaluation shows that querying data from recent years is likely faster than querying older data.

## 8 Related Work

With more devices generating trajectory data [33], systems handling trajectory data must scale. As the amount of data increases, these systems must implement efficient algorithms to query the trajectory data.

Efficiently processing and storing trajectory data is an active research area, with multiple systems used in the industry [22]. These systems differ in how they handle trajectory data. Some use a DW approach [8, 28], and others use a more general trajectory management approach [9, 22, 23, 35].

The underlying storage system varies between using an RDBMS [28], NoSQL databases [22, 23], and processing frameworks based on in-memory storage [35] or distributed file systems [9].

DIPAAL uses the trajectory-based storage model, storing the full trajectories as a single row. Other systems that utilize or support the trajectory-based storage model include Lan et al. [22], Nielsen et al. [28], and Li et al. [23]. Additional storage models are also available, such as the point-based model [22, 28], which stores each point of a trajectory individually, and the segment-based model [9, 22, 35], which stores trajectories in segments. The trajectory-based model is chosen for DIPAAL as it incurs no reconstructing cost during query execution and has reduced storage cost compared to the point-based model [22]. step as no reconstruction is needed.

DIPAAL uses spatial distribution to reduce the runtime when querying data based on the underlying cell grid using Citus shards to distribute and parallelize queries. DIPAAL opted to create the spatial divisions utilizing a kd-tree approach from one year's worth of data, whereas similar systems adopted other approaches, by similar systems [7, 24]. Li et al. [24] chose to use a quad-tree approach to construct their divisions, as they divide non-point data, which is more difficult to find kd-splits for [24]. Aji et al. [7], instead use a custom grid division approach based on thresholds. The spatial distribution approach used in DIPAAL can result in divisions spanning very large areas. However, compared to Aji et al., it is ensured that the largest divisions are split first, and all divisions contain data, enabling a better balance of data across the cluster.

Both Li et al. and Aji et al. use global indices to look up shards. DIPAAL relies on the shards' local indices to quickly evaluate whether a shard contains relevant data.

Various heatmaps are generated as part of DIPAAL's ETL process. DIPAAL utilizes a binning approach similar to Liu et al. [25] by creating small heatmaps based on the 5km cell grid and spatial divisions, which can then be combined into various heatmaps.

## 9 Conclusion

The DIPAAL platform consists of a modular ETL pipeline that enhances the data quality through well-defined cleaning rules. The ETL process's cleaning rules and modularity enables DIPAAL to be applied to any AIS dataset with minimal changes. DIPAAL runs on a cluster of six commodity machines, with the data preparation part of the ETL process capable of running independently.

| | | | Five-worker | | | | | | One-worker | |
|---|---|---|---|---|---|---|---|---|---|---|
| | | | Row | | Columnar | | Columnar partitioned | | Row | |
| | | | Time (s) | WIF (%) | Time (s) | WIF (%) | Time (s) | WIF (%) | Time (s) | Scale Up (%) |
| Aarhus Harbour | 30 day | 200m | 0.37 | 27 | **0.35** | 64 | 0.35 | 47 | 0.59 | 160 |
| | | 50m | 0.36 | 65 | 0.34 | 66 | **0.34** | 53 | 0.55 | 150 |
| | 1 year | 200m | **0.92** | 56 | 0.94 | 80 | 1.89 | 35 | 6.37 | 693 |
| | | 50m | 1.10 | 68 | **1.01** | 82 | 2.83 | 42 | 4.69 | 424 |
| The Great Belt | 30 day | 200m | **0.36** | 21 | 0.47 | 22 | 0.36 | 19 | 0.93 | 258 |
| | | 50m | **1.17** | 39 | 1.36 | 37 | 1.28 | 43 | 1.80 | 154 |
| | 1 year | 200m | **2.19** | 20 | 3.36 | 27 | 2.53 | 23 | 7.68 | 350 |
| | | 50m | **12.50** | 37 | 14.4 | 38 | 13.74 | 42 | 18.17 | 145 |
| Spatial Domain of DIPAAL | 30 day | 200m | **1.27** | 19 | 4.17 | 18 | 1.58 | 15 | 8.09 | 639 |
| | | 50m | **11.90** | 13 | 14.4 | 14 | 12.06 | 14 | 42.10 | 354 |
| | 1 year | 200m | **9.18** | 9 | 35.58 | 16 | 9.36 | 12 | 40.19 | 438 |
| | | 50m | **102.35** | 11 | 126.37 | 10 | 102.81 | 10 | Out of memory | |

Table 1: Heatmap Benchmark Results.

| | | 2011 | | 2021 | | 2022 | |
|---|---|---|---|---|---|---|---|
| | | SD | CV | SD | CV | SD | CV |
| Cell | 50m | 3359k | 66% | 2720k | 50% | 2734k | 51% |
| | 200m | 836k | 65% | 671k | 49% | 6756k | 50% |
| | 1000m | 164k | 61% | 127k | 43% | 129k | 45% |
| | 5000m | 36k | 56% | 32k | 41% | 34k | 44% |
| Heatmap | 50m | 2388MB | 77% | 2002MB | 61% | 1997MB | 61% |
| | 200m | 164MB | 73% | 136MB | 56% | 137MB | 57% |
| | 1000m | 18MB | 56% | 14MB | 40% | 15MB | 42% |
| | 5000m | 11MB | 52% | 9MB | 38% | 9MB | 40% |

Table 2: Spatial Distribution Skewness

Evaluation of the spatial distribution shows that static divisions result in the balance between shards becoming skewed over time.

## Acknowledgement

The paper has received funding from the European Union's funded Project MobiSpaces under grant agreement no 101070279.